\documentclass[twocolumn,showpacs,preprintnumbers,amsmath,amssymb]{revtex4}

\usepackage{graphicx}
\usepackage{dcolumn}
\usepackage{bm}
\usepackage{upgreek}

\begin{document}


\title{Kinetic-inductance-limited reset time of superconducting nanowire photon counters}

\author{Andrew J. Kerman, Eric A. Dauler, and William E. Keicher}
\affiliation{Lincoln Laboratory, Massachusetts Institute of Technology, Lexington, MA, 02420}

\author{Joel K. W. Yang and Karl K. Berggren}
\affiliation{Massachusetts Institute of Technology, Cambridge, MA, 02139}

\author{G. Gol'tsman and B. Voronov}
\affiliation{Moscow State Pedagogical University, Moscow 119345, Russia}

\date{\today}

\begin{abstract}
We investigate the recovery of superconducting NbN-nanowire photon counters after detection of an optical pulse at a wavelength of 1550 nm, and present a model that quantitatively accounts for our observations. The reset time is found to be limited by the large kinetic inductance of these nanowires, which forces a tradeoff between counting rate and either detection efficiency or active area. Devices of usable size and high detection efficiency are found to have reset times orders of magnitude longer than their intrinsic photoresponse time.
\end{abstract}

\pacs{74.76.Db, 85.25.-j}
\maketitle

High-speed photon-counting detectors have many applications, including optical communications \cite{optcomm}, quantum information \cite{quant}, biological physics \cite{bio}, semiconductor processing \cite{semicond}, and laser radar \cite{ladar}. Of particular interest would be a detector that combines ultrafast count rates ($\ge$ GHz) with high single-photon detection efficiency at near-infrared wavelengths; however, current near-infrared photon-counting technologies such as avalanche photodiodes \cite{apds} and photomultiplier tubes \cite{pmts} are limited to much lower count rates by long reset times.

A promising detector technology was reported recently, in which ultrathin superconducting NbN wires are biased with a DC current $I_\mathrm{bias}$ slightly below the critical value $I_\mathrm{C}$ \cite{newtech}. An incident photon of sufficient energy can produce a resistive ``hotspot'' which in turn disrupts the superconductivity across the wire, resulting in a voltage pulse. Observations of this photoresponse showed promise for high counting rates, with measured intrinsic response times as low as $\sim$30 ps \cite{eosample}, and counting rates in the GHz regime \cite{sobrec,newsob}. In this Letter, we present our own investigation into the counting-rate limitation of these devices, in which we directly observe the recovery of the detection efficiency as the device resets (after a detection event), and develop a quantitative model of this process. We find that detectors having both high detection efficiency and usable active area are limited to much lower count rates than studies of their intrinsic response time had suggested \cite{eosample}.

\begin{figure}
\includegraphics[width=3.25in]{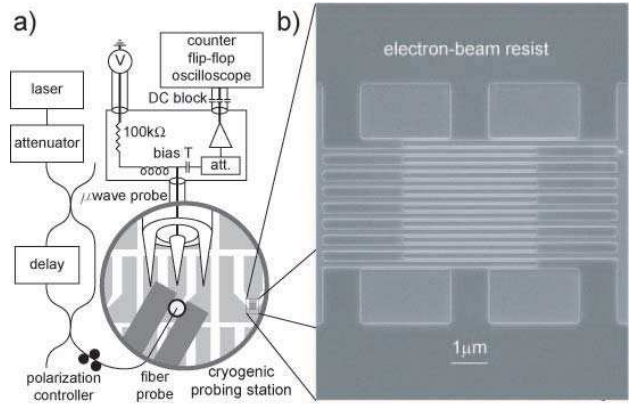}
\caption{\label{fig:schematic} Schematic of the experimental setup. (a) A cryogenic probing station allowed us to probe our devices both electrically and optically. Our samples were $3-8$ mm in size, and typically contained from $50-150$ individual, separately addressable detectors. (b) a scanning electron microscope image of the electron-beam resist pattern used to define a large-area ``meander" structure \cite{newtech,sobrec,newsob}.}
\end{figure}

We fabricated our nanowires using a newly developed process \cite{joel}, on ultrathin ($3-5$ nm) NbN films \cite{films}. We used several geometries, including straight nanowires having widths from $20-400$ nm and lengths from $0.5-50$ \textrm{$\upmu$}m, as well as large-area ``meander" structures \cite{newtech,sobrec} (e.g., Fig. \ref{fig:schematic}(b)) having active-area aspect ratios from $1-50$, fill factors from $25-50$\%, and sizes up to 10-$\upmu$m square. The devices had critical temperatures $T_\mathrm{C}\sim$ $9-10$ K, and critical current densities $J_\mathrm{C}\sim 2-5\times 10^{10}$ A/m$^2$. In total, $\sim$ 400 devices were tested, spanning several different fabrication runs.

\begin{figure}
\includegraphics[width=3.2in]{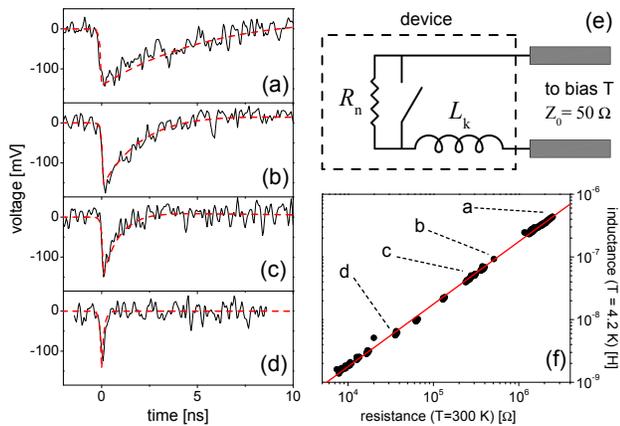}
\caption{\label{fig:pulses} (color online) Inductance-limited recovery of NbN-nanowires. Output pulses are shown for 100 nm wide wires at $T=4.2$ K, with $I_\mathrm{bias}=11.5\;\upmu$A, and dimensions: (a) 10 $\upmu$m $\times$ 10 $\upmu$m meander (total length 500 $\upmu$m); (b) 4 $\upmu$m $\times$ 6 $\upmu$m (120 $\upmu$m); (c) 3 $\upmu$m $\times$ 3.3 $\upmu$m (50 $\upmu$m); (d) 5 $\upmu$m-long single wire. Red dotted lines show the predicted pulse recovery, with no free parameters, for each device based on its measured inductance: $L_\mathrm{k}=$ 415 nH, 110 nH, 44.5 nH, 6.10 nH. These predictions include the effect of the measured $f_\mathrm{L}=15$ MHz and $f_\mathrm{H}=4$ GHz corner frequencies of our amplifiers, and the assumptions: $I_\mathrm{ret}\ll I_\mathrm{bias}$, $R_\mathrm{n}\gg 2\pi f_\mathrm{H}L_\mathrm{k}$;  (e) electrical model; photon absorption corresponds to the switch opening; (f) inductance at $T=4.2$ K vs. room-temperature resistance for 290 individual nanowires from $0.5 - 500$ $\upmu$m long and $20-400$ nm wide, with both straight and ``meander" geometries, from two separate fabrication runs. Points corresponding to the devices of (a)-(d) are indicated. The slope of a linear fit constrained to pass through the origin, shown by a solid line, is $0.997\pm0.002$; this indicates that $L\propto R_\mathrm{300K}$, and therefore that $L$ is predominantly a kinetic inductance.}
\end{figure}

The devices were cooled to as low as 2 K inside a cryogenic probing station (Desert Cryogenics), as illustrated in Fig. \ref{fig:schematic}(a). Electrical contact was established using a cooled 50 $\Omega$ microwave probe (67 GHz bandwidth) attached to a micromanipulator, and connected via coaxial cable to the room-temperature electronics. Current bias was supplied through the DC port of a bias T (Picosecond Pulse Labs 5575A) using a battery-powered, adjustable voltage reference in series with a 100 k$\Omega$ resistor. The AC port of the bias T was connected through a 3 dB attenuator \cite{impedance} to two cascaded low-noise amplifiers (Miteq JS2-00100400-10-10A), and then through a DC block \cite{isuppress} (Inmet 8039) to a 6 GHz real-time oscilloscope (LeCroy Wavemaster 8600A), pulse counter (SRS SR400), and a fast flip-flop (NEL NLG4108) as described below. To optically probe the devices, we used a 1550 nm modelocked fiber laser (Calmar Optcom), with a 10 MHz pulse repetition rate and $\le$1 ps pulse duration, that was sent through an attenuator and polarization controller and then into the probing station via an optical fiber. The devices were illuminated through a lens attached to the end of the fiber (Oz Optics) and mounted to a second micromanipulator arm, at an incidence angle of $\sim 15^\mathrm{o}$ (from normal to the sample surface). The focal spot had a measured e$^{-2}$ radius of $\sim$ 8 $\upmu$m. Figures \ref{fig:pulses}(a)-(d) show output pulses for wires with total lengths from $5-500$ $\upmu$m. The pulses were asymmetric, and longer in duration for longer or narrower wires \cite{dark}.

The microscopic mechanism for the formation and growth of a resistive hotspot after photon absorption has been discussed by other authors \cite{hotspot}. Here, we use a simple phenomenological model, illustrated in Fig. \ref{fig:pulses}(e). A central feature of this model is the kinetic inductance of the wire $L_\mathrm{k}$, which can be much larger than the geometric (magnetic) inductance for very thin films \cite{kadin}. Absorption of a photon corresponds to the switch opening, at which time the wire acquires a resistance $R_\mathrm{n}$ \cite{indsame}. The current in the device then begins to decay from its initial value $I_\mathrm{bias}$ with a time constant $\tau_\mathrm{fall}=L_\mathrm{k}/[50\Omega+R_\mathrm{n}]$, towards a final value $I_\mathrm{n}=I_\mathrm{bias}\times 50\Omega/[50\Omega+R_\mathrm{n}]$. This decay is interrupted, however, at some ``return" current $I_\mathrm{ret}$ when the self-heating of the wire, given by $I(t)^2R_\mathrm{n}$, is sufficiently reduced that the wire becomes fully superconducting again \cite{kadin}. The switch in our model then closes, and the current recovers to its original value with the time constant $\tau_\mathrm{rise}=L_\mathrm{k}/50 \Omega$ \cite{rtime}. From the observed asymmetry of the electrical pulses  ($\tau_\mathrm{fall}\ll\tau_\mathrm{rise}$) \cite{bandwidth}, we conclude that $R_\mathrm{n}\gg 50\Omega$; in this limit, the pulse amplitude reduces to: $V_\mathrm{pulse}\approx(I_\mathrm{bias}-I_\mathrm{ret})\times 50 \Omega\times G_\mathrm{sig}$, where $G_\mathrm{sig}=47.8$ dB is the measured total gain of our signal path in the amplifier passband. Since the observed pulse amplitudes for all devices were well-described by $V_\mathrm{pulse}\approx I_\mathrm{bias}\times 50 \Omega\times G_\mathrm{sig}$, we conclude that $I_\mathrm{ret}\ll I_\mathrm{bias}$. The kinetic inductance $L_\mathrm{k}$, which determines the current recovery time, was measured for each of our devices by fitting the observed frequency-dependent phase shift of a reflected microwave signal. The dashed red lines shown in Fig. \ref{fig:pulses}(a)-(d) are the resulting pulse shapes predicted by our model with no free parameters.

To verify that the large observed inductances were indeed primarily kinetic in nature, we compared them to the corresponding room-temperature resistances $R_\mathrm{300K}$. Kinetic inductance should be proportional to $R_\mathrm{300K}$, since both have the same dependence on wire geometry: $R_\mathrm{300K}=\mathcal{R}_\mathrm{300K}\int ds/A(s)$ and $L_\mathrm{k}=\mathcal{L}_\mathrm{k}\int ds/A(s)$, where $\mathcal{R}_\mathrm{300K}$ and $\mathcal{L}_\mathrm{k}$ are the resistivity and kinetic inductivity, $A$ is the cross-sectional area, and integration is along the wire. This proportionality is demonstrated in Fig. \ref{fig:pulses}(f) for 290 different devices spanning nearly three decades of inductance \cite{lambda}.

To determine the detection efficiency of our devices, we measured the fraction of incident photons that resulted in an output voltage pulse, using a pulse counter \cite{gating}. We varied the discriminator threshold of the counter to identify the voltage range over which the count rate was observed to be constant, and then set the threshold to the center of this range. Varying the polarization of the incident light produced up to a factor of two change in the count rate, and we chose the setting that produced the maximum value. The optical intensity was then chosen such that the optical pulse detection probability was much less than unity. To calibrate the power that the optical probe delivered to each device, we first measured the total optical power exiting the probe with a power meter at room temperature. Next, the peak fraction of this total power subtended by the active area of each device was individually calibrated at low temperature; we scanned the optical probe spatially over each device while recording the count rate, and fit the resulting profiles to the expected convolution of the gaussian-beam and device shape. The resulting (peak) fractions, from $3-80$\% for our meander devices, were then used to calculate detection efficiencies from the observed count rates. The 4 $\upmu$m $\times$ 6 $\upmu$m meander device with 100 nm wire width and 50\% fill factor used in the experiments described below had a measured detection efficiency at 1550 nm, with $I_\mathrm{bias} = 0.98I_\mathrm{C}$, of 2.8\% at 4.2 K, and 5.2\% at 2.1 K.

\begin{figure}
\includegraphics[width=3.2in]{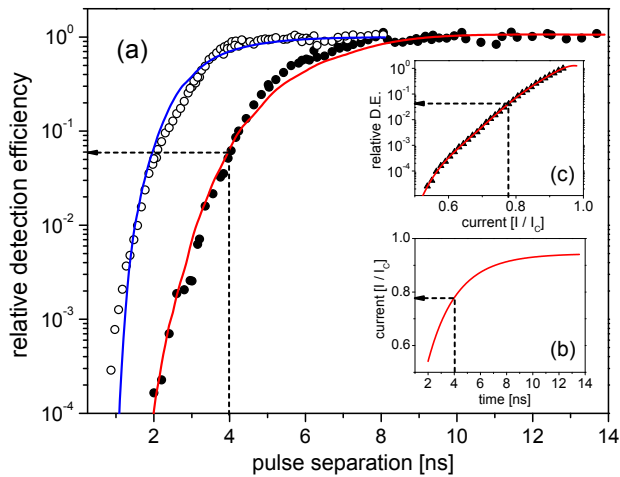}
\caption{\label{fig:doublepulse} (color online) Recovery of the detection efficiency after a detection event. (a) Filled circles are data obtained with a 4 $\upmu$m $\times$ 6 $\upmu$m meander with 100 nm wide wire and a 50\% fill factor, open circles indicate a device of the same size but with a 25\% fill factor. Solid curves are the predictions of our model with no free parameters, obtained as illustrated in (b) and (c) for the 50\% fill device, and based on the measured inductances (109 nH and 47.1 nH at $I_\mathrm{bias}=0$ \cite{current}) and detection efficiency vs. $I_\mathrm{bias}$ (see text).}
\end{figure}

To investigate the time-dependence of the detection efficiency after a detection event, we illuminated the devices with optical pulse pairs \cite{sobrec}, and measured the probability that \textit{both} pulses were detected, as a function of the pulse separation. As shown in Fig. \ref{fig:schematic}, we split the output of our laser into two components, one of which passed through a 0 - 15 ns optical delay line. The two components were then recombined to produce pulse pairs with controllable separation. The output of the amplifiers was sent to a flip-flop, which switches its digital state every time it is triggered \cite{flipthresh}. We then used the oscilloscope to count only those digital pulses from the flip-flop having a width nearly equal to the optical pulse separation, from which we obtained the probability that the device detected both optical pulses in a pair. The result is shown in Fig. \ref{fig:doublepulse} for two different devices, with each scaled to its asymptotic value \cite{intensity}.

The solid lines are the predictions of our model for the two devices, with no free parameters. Figs. \ref{fig:doublepulse}(b)-(c) illustrate how these curves were generated. For a given time on the abscissa of (a), we first found the instantaneous current predicted by our model, based on the measured inductance \cite{current}, as shown in (b). The current was assumed to start at zero (based on our earlier conclusion that $I_\mathrm{ret}\ll I_\mathrm{bias}$) \cite{offset}. The current at each time point was then mapped to a relative detection efficiency (RDE) value using a polynomial fit to the measured detection efficiency vs. $I_\mathrm{bias}$, shown in (c). An example of this mapping is illustrated by the dotted arrows, with $T=$ 4 ns $\rightarrow I/I_\mathrm{C} = 0.78\rightarrow$ RDE $=6\times 10^{-2}$. The resulting predictions agree well with our data for these two devices, which have very different inductances, supporting our model of the reset process in these nanowires.

Fig. \ref{fig:doublepulse} has important implications for high-speed applications of these devices. For the 50$\%$ fill device, it took 8.5 ns for the detection efficiency to recover to 90$\%$ of its initial value, and the device will therefore not support counting rates $\gtrsim$ 120 MHz near full detection efficiency. Although much higher counting rates can readily be achieved with lower-inductance devices, reducing the inductance presently requires either reducing the device area, increasing the wire width, or increasing the film thickness (these likely explain the GHz counting rates observed in \cite{sobrec}), any of which would reduce the system detection efficiency.

In summary, we have shown that the reset time of superconducting NbN-nanowire photon counters is limited by the large kinetic inductance inherent in any thin superconducting film. This result implies that present devices with usable active area and high detection efficiency are intrinsically limited to counting rates well below the GHz regime suggested by early measurements \cite{eosample}, and that any future attempts to increase the counting rates accessible to these devices will have to circumvent their large kinetic inductance. If this can be achieved, the full potential of these devices may become accessible: with an intrinsic photoresponse time at 2 K of only $\sim$30 ps \cite{eosample}, they could extend photon counting into the tens of GHz regime characteristic of modern telecommunications.

We acknowledge D. Oates and W. Oliver (MIT Lincoln Laboratory), S.W. Nam, A. Miller, and R. Hadfield  (NIST) and R. Sobolewski, A. Pearlman, and A. Verevkin (University of Rochester) for helpful discussions and technical assistance. This work made use of MIT's shared scanning-electron-beam-lithography facility in the Research Laboratory of Electronics.

This work is sponsored by the United States Air Force under Air Force Contract \#FA8721-05-C-0002.  Opinions, interpretations, recommendations and conclusions are those of the authors and are not necessarily endorsed by the United States Government.


\begin{thebibliography}{}
\bibitem{isuppress} Without a DC block on the outer conductor, $I_\mathrm{C}$ was suppressed by up to 20\%, probably due to ground noise.
\bibitem{dark} The shape and height of photon-induced counts were observed to be identical to those of spontaneously occurring dark counts, and independent of optical intensity.
\bibitem{indsame} We assumed that the inductance was unaffected by the presence of $R_\mathrm{n}$ (i.e. the normal domain covered only a small fraction of the wire).
\bibitem{rtime} In reality, $I(t)$ and $R(t)$ interact dynamically through Joule heating $I(t)^2R(t)$. Our simplified model approximates this time-dependent $R(t)$ with a square pulse of height $R_\mathrm{n}$, lasting from the photon absorption at $t=0$ to the time when $I(t)=I_\mathrm{ret}$.
\bibitem{bandwidth} The 4 GHz bandwidth of our amplifiers obscured the asymmetry of the pulses from very short wires.
\bibitem{lambda} From this data we can extract the magnetic penetration depth $\lambda$, using: $\mathcal{L}_\mathrm{k}\approx\upmu_0\lambda^2$ (valid in the limit $\lambda\gg t$ where $t$ is the film thickness \cite{kadin}). Assuming $t=4$ nm, we obtain from the inductance data: $\langle\mathcal{L}_\mathrm{k}\rangle =360$ pH$\cdot$nm and $\langle\lambda\rangle = 535$ nm (brackets indicate an average over devices).
\bibitem{gating} The sync output of the laser source was used to trigger the gate of the counter, which we then set to accept signal edges only during a 5 ns window centered on the correct electrical pulse arrival time; this gating allowed us to exclude virtually all dark counts and afterpulsing from the observed count rate.
\bibitem{flipthresh} Using a bias T, the DC offset of the amplified detector signal was shifted relative to the digital switching threshold of the flip-flop input such that the latter was just outside the amplifier noise. This setting gave a digital switching rate for unpaired 10 MHz optical pulses identical to the count rate seen on the pulse counter.
\bibitem{intensity} In order to acquire detection efficiencies over a wide dynamic range in a reasonable time, a relatively high optical intensity was used for these pulse-pair measurements, corresponding to an individual (isolated) pulse detection probability $P_\mathrm{p}\sim$0.57 for both devices (or $N\sim$70 and $N\sim$220 photons per pulse incident on the active areas of the 50\% and 25\% fill factor devices, respectively). Measured values of $P_\mathrm{p}$ were converted to single-photon detection efficiency $P_{1\gamma}$ using the relation: $P_\mathrm{p}=1-\mathrm{exp}(-NP_{1\gamma})$.
\bibitem{current} The inductance was also measured as a function of $I_\mathrm{bias}$, and was observed to increase by up to 20\% close to $I_\textrm{C}$. This was included numerically in our model.
\bibitem {offset} A small time offset was added to each prediction, to account for the time it took the current to decay to zero and the device to return to the superconducting state. These offsets were fixed at the observed fall times of the electrical pulses for each device, 330 and 160 ps for the 50\% and 25\% fill devices, respectively.
\end{thebibliography}
\end{document}